
\font\titlefont = cmr10 scaled \magstep2
\magnification=\magstep1
\vsize=22truecm
\voffset=1.75truecm
\hsize=15truecm
\hoffset=0.95truecm
\baselineskip=20pt

\settabs 18 \columns

\def\b{\smallskip}
\def\bb{\bigskip\bigskip}

\def\ce{\centerline}

\def\no{\noindent}

\ce{\titlefont { An Automatic Invisible Axion }}
\ce{\titlefont{ In The SUSY Preon Model }}

\bb
\ce{{\bf K.S. Babu}$^a$, {\bf Kiwoon Choi}$^b$, {\bf J.C. Pati}$^c$ and
{\bf X. Zhang}$^d$}
\b
\ce {$^a$ Bartol Research Institute}
\ce {University of Delaware, Newark, DE 19716}
\b
\ce {$^b$ Department of Physics}
\ce{Korea Advanced Institute of Science and Technology}
\ce{373-1 Kusong-dong, Yusong-gu, Taejon 305-701, Korea}
\b
\ce {$^c$ Department of Physics}
\ce {University of Maryland,
College Park, MD 20742}
\b
\ce{$^d$Department of Physics}
\ce{Iowa State University, Ames, IA 50011}
\b

\ce{\bf ABSTRACT}
\b
It is shown that the recently proposed
preon model which provides a unified origin
of the diverse mass scales and an explanation of family replication
as well as of inter--family mass--hierarchy, naturally
possesses a Peccei--Quinn (PQ)
symmetry whose spontaneous breaking leads to
an automatic invisible axion.  Existence of the PQ--symmetry
is simply a consequence of supersymmetry and  the requirement of
{\it minimality} in the field--content and interactions,
which proposes that the lagrangian should possess
only those terms which are dictated by the
gauge principle and no others.  In addition to the axion,
the model also generates two superlight
Goldstone bosons and their superpartners all of which are cosmologically
safe.

\bb
\filbreak
{\bf 1.} The idea of the Peccei-Quinn (PQ)
symmetry[1] provides one of the most elegant
solutions to the strong CP problem, in that it forces $\overline{\Theta}
$, determined to be $\le 10^{-9}$, to vanish identically (barring
negligible corrections from the electroweak sector).
In practice,
owing to constraints based on laboratory and astrophysical observations,
the implementation of this idea requires a
light invisible axion[2]. This in turn necessitates
the scale of PQ symmetry breaking
$f_a$
to be rather high: $10^{10} {\rm GeV} \leq f_a \leq 10^{12}{\rm GeV}$[3].

There have been many suggestions in this regard, but invariably extra
 fields and/or the scale of PQ symmetry breaking have to be postulated
from the beginning just to implement this idea[4]. For example, the idea
of the composite axion, suggested by Kim [5,6],
introduces new fermions together with a new axicolor force
which seem to play no other role except implementing
PQ symmetry
and the invisible axion.

The purpose of this note is to point out that a composite invisible axion of
the type envisaged in ref.[5,6]
can naturally occur in
the SUSY preon model[7]
which has evolved over the last few years.
The model exhibits several
attractive features including
an understanding of the origins of (i) diverse mass scales[7],
(ii) protection of the masses of composite quarks and
leptons compared to the scale
of compositeness[8], (iii) family-replication[9] and (iv) inter-family
mass--hierarchy[10].
The model also shows the possibility of a unification of forces
at the level of preons near the Planck scale[11] and
makes many testable predictions[12].
As we will show, the Peccei-Quinn symmetry emerges as an automatic feature
of the model, stemming simply from the requirement of minimality in
field content and interactions on which the model is built.
This amounts to retaining only those terms in the lagrangian which have
purely a gauge origin and no others.
The fermions which make the composite invisible axion are the
preons, which also make quarks, leptons and Higgs bosons.
In other words, they do not have to be postulated just to introduce
the PQ symmetry. Furthermore, the PQ symmetry breaking scale is naturally
identified in the model with the scale $\Lambda_M$ of the preonic metacolor
 force, which represents the scale of compositeness.
 For several independent reasons, including (a) the
 observed mass of
$m_W$[7] (b) the desired unity of forces[11] and (c) a desirable pattern
for the neutrino masses[13], $\Lambda_M$ is determined within the model to lie
around $10^{11}$ GeV, which is precisely
 within the allowed range of $f_a$.

{\bf 2.} To see the origin of the PQ symmetry and that of the invisible axion
 in the model, we need to present some of its salient features.
The model[7] assumes that the effective lagrangian just below the
Planck
scale possesses $N = 1$ local supersymmetry and a gauge symmetry of the form
$G_M
\times G_{fc}$.  Here $G_M = SU(N)_M$ (or SO(N)$_{\rm M}$) generates an
asymptotically free metacolor gauge force which becomes strong at
$\Lambda_M \sim 10^{11}~GeV$ and binds preons.
$G_{fc}$
denotes the flavor-color gauge symmetry, which is assumed to be either
${\cal
G}_{224} = SU(2)_L \times SU(2)_R \times SU(4)^c$[14], or a subgroup of
${\cal
G}_{224}$ containing SU(2)$_{\rm L} \times$ U(1)$_{\rm Y} \times$ SU(3)$^{\rm
c}$.
The gauge symmetry $G_M \times G_{fc}$ operates on a set of six positive
and six
negative {\it massless} chiral preonic superfields $\Phi^{a,\sigma}_{\pm} =
(\varphi, \psi, F)^{a,\sigma}_{L,R}$, each belonging to the fundamental
representation $N$ of $SU(N)_M$.  Thus ``a'' runs over {\it six values}:
\ $(x,y);
(r,y,b,\ell)$, where $(x,y)$ denote the two basic flavor-attributes $(u,d)$
and
$(r,y,b,\ell)$ the four basic color-attributes of a quark-lepton
family[14].  The index $\sigma$ runs over metacolor quantum numbers.
The representation
content
of the preonic superfields under the gauge symmetry is shown below:
$$
\matrix{ &  & SU(2)_L \times & SU(2)_R  \times & SU(4)^C_{L+R}  \times &
SU(N)_{L+R} \cr
\Phi_+^{f,\sigma} = (\varphi^f_L, \psi^f_L, F^f_L)^{\sigma} & \sim &
2_L, &
1, & 1, & N \cr
\Phi_-^{f,\sigma} = (\varphi^f_R, \psi^f_R, F^f_R)^{\sigma} & \sim & 1,
&
2_R, & 1, & N \cr
\Phi_+^{c,\sigma} = (\varphi^c_L, \psi^c_L, F^c_L)^{\sigma} & \sim & 1,
& 1,
&  4_c, & N \cr
\Phi_-^{c,\sigma} = (\varphi^c_R, \psi^c_R, F^c_R)^{\sigma} & \sim & 1,
& 1,
& 4_c, & N  \cr}$$
Here $f$ stands for two preonic flavors $\equiv (u, d)$, while $c$ denotes four
colors ($r,
y, b$, and $\ell$).  {\it Note that there is no repetition of any entity at the
preon level}.

To be specific, we will assume that, just below the Planck scale,
the gauge symmetry is
$G_{gauge}= SU(6)_M \times G_{fc}$
where $G_{fc} = G_{224}$,
although our conclusion
will not alter
if $G_{f c}$ is a subgroup of
${\cal G}( 2, 2, 4)$ containing the standard model[15].

In the interest of minimizing or removing all arbitrary parameters, the
preon model[7] adheres to a principle of minimality, which proposes that
the field content must be minimal, with no repetition of preonic
entities, and so also must be the interactions.  {\it Such a principle
requires that the lagrangian should possess only those
terms which are dictated by the gauge--covariant derivatives
and no others}.  This in
particular permits no non--gauge mass, Yukawa and quartic couplings,
which incidentally have been the main source of arbitrariness in the
conventional approach to unification,
based on fundamental Higgs bosons, quarks and
leptons.  By contrast, the preon model introduces no more than a few
gauge couplings (e.g. $\alpha_M$, $\alpha_2$ and $\alpha_4$, defined
near the Planck scale) as its {\it only parameters}.
(Even these few gauge couplings would merge into
one, if there is an underlying unity of forces at the preon--level, near
the Planck scale[11]).  Yet, it has been shown to be viable and capable
of addressing some major issues[7-11].

The lagrangian of the preon model[7], restricted by such a principle,
consists only of the minimal gauge interactions and possesses $N=1$
local SUSY.  In other words, it has no $F$--term.  Given six preonic
attributes
it is easy to verify that in the absence of the flavor--color gauge
interactions (generated by $G_{224}$), such a lagrangian would possess
a global symmetry $SU(6)_L \times SU(6)_R \times U(1)_V \times U(1)_X$,
while in their presence the global and local symmetry of the model,
ignoring color anomaly for a
moment, is given by:
$$\eqalign{
G_P =  &  {[ U(1)_V \times U(1)_X \times U(1)_{T_L}
\times U(1)_{T_R} ]}_{( global )} \cr
 & \times {[ SU(6)_M \times SU(2)_L \times SU(2)_R \times SU(4)^c
                         ]}_{(local)} \cr} \eqno(1)
$$

\noindent $U(1)_V ~~{\rm and}~~ U(1)_X$
denote the preon-number and the non-anomalous R-symmetry
respectively[16],
which in our case assign the following charges to the preons:
$$
Q_V( \Phi_{L,R}^{a, \alpha} ) = 1;~~~~
Q_X( \varphi_{L, R}^{a, \alpha}  )  =0 $$
$$
Q_X(\Psi_L^{(a,\alpha)}) = -Q_X(\Psi_R^{(a,\alpha)}) = -Q_X(\lambda) = 1
\eqno(2)
$$
(Here the superspace Grassmann coordinate has $N_X=1$.)
$T_{L, R}$ have the following representations in the space of
${( r, y, b, l, u, d)}$:
$${
T_{L, R} = {\rm diag.}(1,1,1,1,-2,-2)_{L,R}.}\eqno(3)
$$

\no It is useful to combine the four global $U(1)$'s listed in (1) as
follows:

$$\eqalign{
T_+   =  T_L + T_R;~~~~~~ & Q_V \cr
{}~~~~~~~~~~~~    Q_-  =  Q_X - (T_L - T_R );
{}~~~~~~~  Q_+   = & Q_X + ( T_L - T_R ) ~~. \cr }  \eqno(4)$$

\no Of the
four global $U(1)'s$,  only $Q_{\pm}$ are chiral, among these only
$Q_+$ has a nonzero
SU(3)-color anomaly and
thus it is the one which serves as the Peccei-Quinn charge.

We stress that the existence of the four global U(1) symmetries including
$U(1)_X$, and thus of the PQ symmetry $Q_+$,
is a natural feature of the model, in the sense that they emerge
simply from the assumption of minimality in field content and
interactions[17].
We now discuss spontaneous
breaking of some or all of these symmetries.

{\bf 3}.  $\underline{Symmetry  ~Breaking:}$
It is assumed that as the asymptotically free metacolor force becomes
strong at a scale $\Lambda_M \sim 10^{11}~GeV$, (a) it confines preons
to make composite quarks and leptons, and (b) it forms a few
SUSY--preserving and also SUSY--breaking
condensates, all of which preserve metacolor.  The latter
include the metagaugino pair $<\vec{\lambda}.\vec{\lambda}>$ and the
preonic fermion--pairs $<\overline{\psi}^a\psi^a>$, both of which break
SUSY in a massless preon theory.  Noting that in the model under
consideration, a dynamical breaking of SUSY would be forbidden owing to
the Witten--index theorem[18], it has been argued[8] that each of these
fermionic condensates, which do break SUSY, must be damped by
$(\Lambda_M/M_{Pl})$, so that each would vanish in the absence of
gravity (i.e., as $M_{Pl} \rightarrow \infty$).  We thus expect[8,7]:
$$
<\overline{\psi}^a\psi^a> = a_{\psi_a}\Lambda_M^3(\Lambda_M/M_{Pl});~~~~
<\vec{\lambda}.\vec{\lambda}> = a_\lambda \Lambda_M^3(\Lambda_M/M_{Pl})
\eqno(5)
$$
where $a_{\psi_a}$ and $a_\lambda$ are apriori expected to be of order
unity.  These induce SUSY--breaking mass--splittings
$\delta m_S \sim \Lambda_M (\Lambda_M/M_{Pl}) \sim 1~TeV$.  In addition,
$<\overline{\psi}^a\psi^a>$--condensates break $SU(2) \times U(1)$ for
$a=(x,y)$ and give masses to $W$ and $Z$ of order $(1/10)\Lambda_M(
\Lambda_M/M_{Pl}) \sim 100~GeV$, as well as to quarks and leptons which
are $\le 100~GeV$[7,10].

The SUSY--preserving condensates have no reason to be suppressed.  They
are thus expected to be of order
$\Lambda_M \sim 10^{11}~GeV$.  Although, in principle, the pattern of
condensates which form, should be derivable from the underlying preonic
theory, in practice, one is far from being able to do so.  This is
because of our inexperience in dealing with the non--perturbative
dynamics of SUSY QCD (leaving aside, of course, some general results
like the index theorem[18]).  The preonic idea seems most attractive and
thus worth pursuing, nevertheless, because of its utmost economy in
parameters.  With this in view, we proceed by making a {\it broad dynamical
assumption} which is this:  (a) SUSY QCD (with $m^{(0)}_\psi =
m^{(0)}_\phi = 0$), unrestricted by the constraints of the Vafa-Witten
theorem[19], permits a dynamical breaking of parity and vectorial
symmetries like ``isospin'' ($SU(2)_{L+R}$), baryon and/or lepton numbers;
(b) A suitable set of metacolor--singlet, SUSY--
preserving, condensates form, which break the gauge symmetry $SU(2)_L
\times SU(2)_R \times SU(4)^c$ into the standard model symmetry
$SU(2)_L \times U(1)_Y \times SU(3)^c$, and simultaneously also all or
at least a major subset of the global $U(1)'$s listed in (1), at the
scale $\Lambda_M$[20].  To be specific, it is assumed that the
SUSY--preserving
metacolor singlet condensates (written in a schematic notation)
include:

$$\eqalign{
< \Delta_R > & = < \psi_R^{u \alpha } \psi_R^{u \beta} \phi^{l*}_{L \alpha}
                           \phi_{L \beta}^{l*} > \sim ( 1, 3_R, 10^c )
\cr
< S_R > & = < \phi_R^{u \alpha} \phi_R^{u \beta} \phi_{L \alpha}^{l*}
                      \phi_{L \beta}^{l*} > \sim (1, 3_R, 10^c ) \cr
< \xi_1 > & = < \epsilon_{ \alpha \beta \gamma \delta \rho \sigma}
                 \phi_R^{r \alpha} \phi_R^{y \beta}
                     \phi_R^{b \gamma} \phi_R^{l \delta}
               \phi_R^{u \rho} \phi_R^{d \sigma} > \sim
                 ( 1, 1, 1^c )    \cr
< \xi_2 > & = <  \epsilon_{\alpha \beta \gamma \delta
                \rho \sigma} \phi_R^{r \alpha} \phi_R^{y \beta}
                \phi_R^{b \gamma} \phi_R^{l \delta}
                \phi_L^{u \rho} \phi_L^{d \sigma} > \sim
             ( 1,1, 1^c )  ~~.}   \eqno(6)
$$

\no The transformation property of the full-multiplet,
containing the condensate, under $SU(2)_L \times
  SU(2)_R \times SU(4)^c$ is exhibited on the right.  Antisymmetrisation
on the $SU(3)^c$--indices is to be understood.  While the formation of
these condensates is a dynamical assumption of the model[20], it may be
noted that each of these condensates is at least in a highly attractive
channel.  For example, $\Delta_R$ involves
       ${( \psi \phi^* )}_{Adjoint} ~ {( \psi \phi^* )}_{Adjoint}$
and
$S_R$ involves
${( \phi \phi^* )}_{Adjoint} ~ {( \phi \phi^* )}_{Adjoint}$
combinations. Unlike $< \overline{\psi} \psi > ~~{\rm and}~~<\lambda\lambda>$,
these condensates can form, while preserving SUSY.   Their quantum
numbers are listed below:

$$\matrix{    &  B-L  & I_{3R} & Q_V &  Q_X & T_L & T_R \cr
 < \Delta_R > & -2    &  1     &   0 &  -2  &  -2 &  -4 \cr
 < S_R >      &  -2  &   1     &  0  &  0   &  -2  & -4 \cr
< \xi_1 >     & 0    &   0     &   6 &  0   &  0   & 0 \cr
< \xi_2 >     & 0    &   0     &    6 & 0   &  -4  &  4 \cr } \eqno(7)
$$

It is then easy to see that the condensates of eq. (6) break the symmetry
group $G$ (see eq. (1)) down to
$${
SU(2)_L\times U(1)_Y\times SU(3)^c\times [U(1)_H]_{global}},
\eqno(8)$$
where $Y=I_{3R}+(B-L)/2$ is the weak hypercharge
and the charge for $U(1)_H$, in the space of the preon--flavors
$(r,y,b,l,u,d)_{L,R}$, is given by

$$\eqalign{
Q_H & =T_L+T_R-3(B-L)
\cr &= {\rm diag}.(2,2,2,-2,-2,-2)_{L,R}}\eqno(9)$$

\no The charge $Q_H$ is vectorial and acts
effectively as baryon number, because
$$
Q_H({\rm quarks}~ \sim \psi_u\varphi_r^*) = -4;~~
Q_H({\rm leptons}~\sim \psi_u\varphi_l^*) = 0~~. \eqno(10)
$$

\no If $Q_H$ is preserved, proton will be stable.  It is possible
that $Q_H$ breaks through additional condensates so as to induce proton
decay (see remarks later).  Even without such condensates,
$Q_H$ breaks through $SU(2)_L$-anomaly, however the
proton decay induced by $SU(2)_L$-instantons would
be too slow to be observable, as
observed by 't Hooft[21].



{\bf 4.} {\it Supermultiplets of axion and other Goldstone bosons:}

As mentioned above, of the four global $U(1)$ symmetries listed in (4),
only the
combination $Q_H =T_L+T_R-3(B-L)$
is preserved, while the remaining three --i.e., $Q_V,~Q_-$
and $Q_+$ -- are broken spontaneously
at $\Lambda_M$  by the condensates listed in (6).
Such a breaking thus generates {\it three physical Goldstone bosons}
each with a
decay constant of the order of $\Lambda_M$.

Of the three spontaneously broken global $U(1)$'s, $Q_V$ and $Q_-$ are
broken explicitly by $SU(2)_L$ anomaly, while $Q_+$ is broken by both QCD
and $SU(2)_L$ anomalies.  Thus $Q_+$ can be identified as a PQ symmetry.
 Its existence provides the familiar resolution of the strong--CP
problem[1], and its spontaneous breaking at $\Lambda_M$ generates the
standard invisible axion (see below).

We now observe that although $Q_V$ and $Q_H$ are broken
explicitly by $SU(2)_L$
anomalies, the linear combination $Q_V^\prime \equiv (2Q_V+Q_H) =$
$diag.(4,4,4,0,0,0)_{L,R}$ is free from anomalies.
We therefore expect that there will be one {\it exactly
massless} Goldstone boson associated with the spontaneous breaking of
$Q_V^\prime $.  For this reason, we will discuss the Goldstone boson
spectrum in terms of $Q_+, Q_-$ and $Q_V^\prime$.

We denote the Goldstone bosons corresponding to spontaneous breaking of
$Q_+, Q_-$ and $Q_V^\prime$ by $a_+, a_-$ and $a_V$ respectively, which
are named as follows:
$$
(Q_+,Q_-,Q_V^\prime) \rightarrow \{{\rm axion}(a_+), {\rm chion}(a_-),
{\rm vion}(a_V)\} \eqno(11)
$$
\noindent
Let ``$\phi_i$'' denote these three Goldstone bosons.  At energy scales
well below $\Lambda_M$, the effective interactions of $\phi_i$ can be
written as
$${
{{1}\over{\Lambda_M}}\partial_{\mu}\phi_i J^{\mu}_i+{{1}\over{16\pi^2}}
{{\phi_i}\over{\Lambda_M}}
(c_i G\tilde{G}+d_i W\tilde{W})},\eqno(12)
$$
where $J^{\mu}_i$ denotes the appropriate current mode of generic
composite fields, e.g. quarks and leptons, with masses far below
$\Lambda_M$, $G$ and $W$ are the gluon and $W$--boson field strengths
respectively, and $\tilde{G}$ and $\tilde{W}$ are their duals.  The
coefficients $c_i$ and $d_i$ are in general of order unity, except for
$Q_V^\prime$, for which $c=d=0$.

The axion coupling to QCD anomaly generates an axion mass \hfil\break
$m_a \sim f_\pi m_\pi/\Lambda_M \sim 10^{-4}~eV$, with
$\Lambda_M \sim 10^{11}~GeV$.  This is the invisible axion, which is a
leading candidate for cold dark matter.  As is well known, for
$\Lambda_M \sim 10^{11}~GeV$, axions and other Goldstone bosons, with
the coupling of eq. (12) are consistent with all phenomenological
constraints including those arising from astrophysical and cosmological
arguments.

As for the other two Goldstone bosons -- chion ($a_-$) and vion ($a_V$)
-- $SU(2)_L$--instantons induce a mass for $a_-$, but not for $a_V$,
since $Q_V^\prime$ has no $SU(2)_L$ anomaly.  Suppressed by the
$SU(2)_L$--instanton factor and also the small neutrino masses, the mass
of $a_-$ is expected to be lighter than $(v_{EW}/\Lambda_M)^{1/2}
(v_{EW})e^{-4\pi^2/g_2^2} \sim 10^{-34}~eV$.  If it exists, chion may well be
the {\it lightest massive particle} of nature.  Because of its coupling
to $W \tilde{W}$ (see eq. (12)), it would decay into two photons with a
lifetime far exceeding the age of the universe.  The vion, on the other
hand, is exactly massless and stable.  Such ultralight and very weakly
interacting objects would, of course, have no cosmological significance,
unlike the axion.

Because of supersymmetry, the Goldstone bosons $(a_i)$, $i=+,-,V$ will
be accompanied by spin--1/2 partners $\tilde{a}_i$ and real
spin--0 scalar partners $s_i$, which form the Goldstone supermultiplets:
$A_i = (s_i+ia_i, \tilde{a}_i, F_i)$.  These superpartners corresponding
to $i=+,-,V$ may be named as: (saxion ($s_+$), axino ($\tilde{a}_+)$);
(schion ($s_-$), chino ($\tilde{a}_-$)); and (svion ($s_V$), vino
($\tilde{a}_V$)).

The couplings of $s_i$ and $\tilde{a}_i$ may be obtained from the
supersymmetrized form of (12) which is given by
$${
{{k_{ij}}\over{\Lambda_M}}[(A+\bar{A})\bar{Z}_jZ_j]_D+
{{1}\over{16\pi^2}}[{{A_i}\over{\Lambda_M}}(c_iW_GW_G+d_iW_WW_W)]_F+{\rm h.c}}
 .
\eqno(13)$$
\noindent
Here $Z_j = (z_j, \chi_j,F_j)$
denote   composite chiral matter superfields
including those of quarks and leptons whose possible
gauge interactions are ignored here,
$W_G$ and $W_W$ are the  gauge-covariant chiral gauge superfields for
$SU(3)^c$ and $SU(2)_L$ respectively, and the subscripts
$F$ and $D$ stand for the $F$ and $D$-components of superfields.
The coefficients $k_{ij}$ would
depend upon the details of metacolor dynamics, but are expected to
be of order unity in general since they are $not$ forbidden
by any of the unbroken symmetries.

Clearly the low energy couplings of all three pairs of superpartners --
i.e., (axino, saxion), (chino, schion) and (vino, svion) -- to normal
matter are suppressed
by $1/\Lambda_M$ as those of the axion.
Although these superpartners
will not lead to any significant consequences for accelerator
experiments, since
$\Lambda_M\sim 10^{11}$ GeV,
they may play some role in cosmology.
In the SUSY preon model, it is expected
that each of these superpartners have masses of $10^2-10^3$ GeV,  which
is the effective SUSY--breaking scale.
Then the effective interactions of eq. (13),
which would allow coupling of the form $\tilde{a}_i \rightarrow
\tilde{W}_{1/2}W$ and $s_i \rightarrow W^+W^-$,
would provide
a variety of decay channels of these pairs:
e.g. $\tilde{a}_i\rightarrow z_j\chi_j, \gamma\tilde{\gamma}$, and
$s_i\rightarrow \chi_j\bar{\chi_j},
\gamma\gamma$, and possibly also
$\tilde{a}_i \rightarrow W \tilde{W}_{1/2}$ and $s_i \rightarrow WW$ (if
$\tilde{a}_i$ and $s_i$ are massive enough).
The possible decay modes and
the corresponding life  times $\tau(\tilde{a}_i,s_i)$ depend on the details
of the  mass spectrum and also the mixings among involved
particles.  With their masses around $10^2\sim 10^3$ GeV
(and $\Lambda_M\sim 10^{11}$ GeV), we estimate that
$\tau(\tilde{a}_i,s_i)\ll 1~ sec$.  This is cosmologically safe.



It is worth noting that the effective coupling of the light Goldstone
bosons with the $SU(2)_L$ gauge bosons, induced by anomaly, would in
turn lead to their couplings with fermion--pairs, whose strengths would
be of order $\alpha_W^2m_f/\Lambda_M$; where $m_f$ is the mass of the
relevant fermion.  Since the energy loss in red giants puts a constraint
on the coupling of any such light particle to $e\overline{e}$ pair to be
less than $10^{-13}$[3], it follows that $\Lambda_M \ge 10^{6}~GeV$.
This, of course, is satisfied in the
preon model, since $\Lambda_M \sim 10^{11}~GeV$.

Before closing, a few comments are in order: (1) First,
we have checked that our scheme of
spontaneous symmetry breaking yields three disconnected
degenerate vacua, and thus results in domain walls.  One would
then need inflation to resolve the domain wall problem.  Inflation is,
of course, needed in any model to resolve other cosmological issues, in
particular the horizon and the flatness problems.  The possibility of
generating a satisfactory potential for implementing the ``new''
inflation scenario in a SUSY preon model, because of SUSY--forbiddenness
of certain mass and coupling parameters, has been considered
elsewhere[22].

(2)  Although the discussions in this paper are based on a specific set
of condensates (eq. (6)), our conclusion about the existence of an
invisible axion is more general.  In fact, any alternative set of
condensates which breaks the preonic symmetry $G_P$ to the standard
model gauge symmetry at $\Lambda_M$ will automatically yield the
standard invisible axion.  This is because $\Lambda_M$ is determined
within the model on {\it other grounds} to be about $10^{11}~GeV$.

(3) Third, we note in passing that the condensate pattern listed in (6)
violates lepton number and gives a Majorana mass to the right--handed
neutrinos, but it conserves $Q_H$ and thus leaves proton stable.  Baryon
non--conservation and proton decay would, however, occur at an
observable rate if two additional condensates such as
$<\Sigma>~~ \propto~~ <\psi_L^u\psi_L^d\psi_R^u\psi_R^d\varphi_L^u
\varphi_L^d>$ and $<\zeta>~~ \propto~~ <\varphi_L^r\varphi_L^y\varphi_L^b
\varphi_L^l\varphi_L^u\varphi_L^d>$, which preserve $SU(6)_M$ and
$SU(2)_L \times SU(2)_R \times SU(4)^c$, form.  One can argue that
$<\Sigma>$ breaks SUSY while $<\zeta>$ preserves SUSY.
  Thus, following Ref. 8, one would
expect that $<\Sigma> \sim \Lambda_M (\Lambda_M/M_{Pl})$ and
$<\zeta> \sim \Lambda_M$.  This leads to an amplitude for
$(3q \rightarrow \overline{l})$ of order $(\Lambda_M/M_{Pl})/
\Lambda_M^2 \sim 10^{-8} \times 10^{-22}~GeV^{-2} \sim
10^{-30}~GeV^{-2}$.  Remarkably enough, this is $precisely$ the right
order of magnitude for proton to decay into $e^+\pi^0$ with lifetime
$\sim 10^{32}-10^{33}$ yrs.  This point will be considered in more
detail elsewhere.

To conclude, we see that the preon model, subject to the assumption of
minimality in field content and interactions automatically possesses a
PQ symmetry.  Subject to a broad dynamical assumption about the pattern
of symmetry breaking, it leads naturally
to an invisible axion
because the symmetry breaking scale $\Lambda_M$ is determined on {\it
other grounds} to be about $10^{11}~GeV$.  The model also generates one
massless and one
superlight Goldstone boson with mass $\le 10^{-34}~eV$ as well as the
spin-0 and spin-1/2 superpartners of all three Goldstone bosons with
masses of the order of $10^2-10^3$ GeV.
All of these particles are cosmologically safe, either because they are
superlight and very weakly coupled
 or because they are sufficiently short--lived.  Meanwhile the axion
serves as a strong candidate for cold dark matter.

{\bf Acknowledgement}
The research of K.S.B is supported in part by a grant from the
Department of Energy.
X.Z is
supported in part by the Office of High Energy and Nuclear Physics of the
U.S. Department of Energy (Grant No. DE-FG02-94ER40817).
 J.C.P is supported in part by a grant from the
National Science Foundation.
The work of K. Choi is supported in part
by KOSEF  through the CTP at Seoul National University.
K. Choi and J. C. Pati acknowledge
the hospitality of the ICTP, Trieste  where part of  this work was done.
JCP wishes to thank Michael Dine,
Andre Linde and Helen Quinn for helpful discussions.

\ce {\bf References}

\item{[1]} R.D. Peccei and H. Quinn, Phys. Rev. Lett. {\bf 38}, 1440 (1977).

\item{[2]} J.E. Kim, Phys. Rev. Lett. {\bf 43}, 103 (1979); M.A. Shifman,
A.I. Vainshtein and V.I. Zakharov, Nucl. Phys. {\bf B166}, 199 (1981);
M. Dine, W. Fischler and M. Srednicki, Phys. Lett. {\bf B104}, 199
(1981); A.P. Zhitniskii, Sov. J. Nucl. Phys. {\bf 31}, 260 (1980).

\item{[3]} For a review see, J.E. Kim, Phys. Rept. {\bf 149}, 1 (1987);
M. S. Turner, Phys. Rep. {\bf 197}, 67 (1990); G. G. Raffelt,
Phys. Rep. {\bf 198}, 1 (1990).

\item{[4]} For some notable exceptions, see e.g. P. Langacker, R. Peccei
and T. Yanagida, ~Mod. Phys. Lett. {\bf 1}, 541 (1986); K. Kang and M.
Shin, $ibid$., {\bf 1}, 585 (1986); D. Chang and G. Senjanovic, Phys.
Lett. {\bf B188}, 231 (1987); G. Lazarides, C. Panagiotakopoulos and
Q. Shafi, Phys. Rev. Lett. {\bf 56}, 432 (1986).

\item{[5]} J.E. Kim, Phys. Rev. {\bf D 31}, 1733 (1985).

\item{[6]} K. Choi and J. E. Kim, Phys. Rev. {\bf D 32}, 1828
(1985).

\item{[7]} J.C. Pati, Phys. Lett. {\bf B228}, 228 (1989); For a recent
review, see J.C. Pati, ``Towards a Pure Gauge origin of the Fundamental
Forces: Unity through Preons'', UMD Publication 94-069, Proceedings SUSY
Conf. held at Boston, ed. by P. Nath, (World Scientific) page 255.

\item{[8]} J.C. Pati, M. Cvetic and H. Sharatchandra, Phys. Rev. Lett.
{\bf 58}, 851 (1987).

\item{[9]} K.S. Babu, J.C. Pati and H. Stremnitzer, Phys. Lett. {\bf B256},
206 (1991).

\item{[10]} K.S. Babu, J.C. Pati and H. Stremnitzer, Phys. Rev. Lett.
{\bf 67}, 1688 (1991).

\item{[11]} K.S. Babu and J.C. Pati, Phys. Rev. {\bf D 48}, R1921 (1993).

\item{[12]} See ref.[7,10] and K.S. Babu, J.C. Pati and X. Zhang, Phys. Rev.
{\bf D 46}, 2190 (1992).

\item{[13]} K.S. Babu, J.C. Pati and H. Stremnitzer, Phys. Lett.
{\bf B264}, 347 (1991).

\item{[14]} J.C. Pati and A. Salam, Phys.
Rev. Lett. {\bf 31}, 661 (1973);
Phys. Rev. {\bf D10}, 275 (1974).

\item{[15]} Although much of our results is not tied to a precise choice
of $G_M$, the choice $G_M = SU(6)$ seems to be suggested by the
following set of considerations:  (i) the desire to achieve
unification[11], which suggests that, if $G_M = SU(N)$, $N$ should not
be smaller than 4 and larger than 6; (ii) the need to avoid Witten
$SU(2)$--anomaly which says that $N$ should be even and (iii) the fact
that we utilize the Witten index theorem and that the index has been
calculated and found to be non--zero for the case of massless preons only
when the number of preon flavors which is six, is an integer multiple of
the ``number'' of metacolor $N$ (see the last paper in Ref. 18).

\item{[16]} I. Affleck, M. Dine and N. Seiberg, Nucl. Phys. {\bf 241B},
493 (1984); Nucl. Phys. {\bf 256B}, 557 (1985).  Note that in our case,
the number of preon flavors is equal to the number of metacolors, both
of which are six.  As discussed in the reference mentioned above, in
this case, the instantons do not contribute to the superpotential and
thus do not help choose between $\varphi_{L,R}=0$ (confining solution) and
$\varphi_{L,R} \rightarrow \infty$ (Higgs mode).  Thus, at least there is
no general argument preventing the SUSY preon theory from becoming
strongly interacting and confining, while preserving the metacolor gauge
symmetry corresponding to $\varphi_{L,R}=0$ (see Ref. 8 for further
discussions).

\item{[17]} We note that if one did not insist on the principle of
minimality, one could, of course, have introduced arbitrary mass terms
for the color--carrying preons by allowing terms such as
$m\Phi_+^c\Phi_-^{c*}$ in the superpotential, which would be
gauge--invariant, but which would not respect $U(1)_X$, $T_L$ and
$T_R$.  Thus it is the minimality in the interactions which ensures the
symmetries $U(1)_X, T_L$ and $T_R$.  Conversely, if one imposes $U(1)_X$
in addition to gauge invariance and supersymmetry, $no$ non--gauge terms
(i.e., no $F$--term) would be allowed and
minimality would follow (see also remarks in Ref. 17).
We are aware that Planck scale effects involving quantum
black hole and/or worm hole effects could in general erase global
symmetries like $U(1)_X$ and thus need not respect the principle of
minimality.  (See eg. T. Banks, Physicalia, {\bf 12}, 19 (1990); J.
Preskill, Proceedings of the Int. Symp. of Gravity, the Woodlands, Texas
(1992).)  Since the analysis of these effects, based
on euclidian metric and not well--understood quantum gravity, is still
not conclusive, at least in so far as the magnitude of these effects,
we ignore their implications for the present.  It is
conceivable that one would find additional reasons to either forbid or
suppress sufficiently the unwanted Planck--scale effects anyhow.

\item{[18]} E. Witten, Nucl. Phys. {\bf B 202}, 253 (1982); S. Cecotti
and L. Giradello, Phys. Lett. {\bf B110}, 39 (1983); E. Cohen and L.
Gomez, Phys. Rev. Lett. {\bf 52}, 237 (1984).

\item{[19]} C. Vafa and E. Witten, Phys. Rev. Lett. {\bf 53}, 535 (1984)
and Nucl. Phys. {\bf B234}, 173 (1984).

\item{[20]} To draw a perspective, while the assumed pattern of symmetry
breaking is not altogether implausible in that the needed condensates
are in highly attractive channels, the choice in this regard is
nevertheless arbitrary.  This represents a negative element in the
preonic approach.  On the positive side, while the pattern of
condensates needs to be assumed, the scales of
the condensates including those with a damping by $(\Lambda_M/M_{Pl})$
are motivated on general grounds.  Furthermore, corresponding to the
broad assumption about the pattern of symmetry breaking, the preon model
makes a {\it host of predictions}[7,9,10,12], on the basis of which the
said assumptions can be amply tested.  It also explains the
inter--family mass--hierarchy and the hierarchy of scales from
$M_{Pl}$ to $m_\nu$.  These are among the main reasons why one feels
``justified'' to pursue the preon model, in spite of the needed
dynamical assumption about the pattern of symmetry breaking.
For what it is worth, we remark in passing that a certain arbitrariness,
stemming from ignorance in dynamics, is not exclusive to the preon
theory, but is present in more ambitious approaches as well.  Compare
e.g. the needed VEVs of the scalar $\nu_R$ and $N$ in the
three--generation Calabi--Yau models, or for that matter the needed
assumption that higher dimensional superstring theories compactify into
four dimensions.  For the elementary Higgs--picture, there is, of
course, the complete arbitrariness in the choice of representations of
the Higgs multiplets and the associated parameters, and thereby in the
pattern of VEVs.  The preon model has the relative merit, as mentioned
above, that at least it makes certain crucial predictions, by which it
can be excluded, if it is wrong, or confirmed, if it is right.

\item{[21]} G. 't Hooft, in {\it Recent Developments in Gauge Theories},
eds. G. 't Hooft et. al., (Plenum, NY 1980), p. 135.

\item{[22]} M. Cvetic, T. Hubsch, J.C. Pati and H. Stremnitzer, Phys. Rev.
{\bf D 40}, 1311 (1989).

\bye